\documentclass[twoside,english]{iopart}
\usepackage[T1]{fontenc}
\usepackage[latin9]{inputenc}
\usepackage{geometry}
\usepackage{float}
\usepackage{amsbsy}
\usepackage{graphicx}

\makeatletter

\providecommand{\tabularnewline}{\\}

\usepackage{iopams}
\usepackage{setstack}

\usepackage{babel}

\makeatother

\begin{document}
\title{Exact densities of loops in O(1) dense loop model and of clusters
in critical percolation on a cylinder II: rotated lattice. }
\author{{\Large{}{}A.M. Povolotsky}}
\address{{\large{}{}Bogoliubov Laboratory of Theoretical Physics, Joint Institute
for Nuclear Research, 141980, Dubna, Russia}\\
 {\large{}{}National Research University Higher School of Economics,
20 Myasnitskaya, 101000, Moscow, Russia}}
\begin{abstract}
This work continues the study started in \cite{Povolotsky2021}, where
the exact densities of loops in the O(1) dense loop model on an infinite
strip of the square lattice with periodic boundary conditions were
obtained. These densities are also equal to the densities of critical
percolation clusters on the forty five degree rotated square lattice rolled
into a cylinder. Here, we extend those results to the square lattice
with a tilt. This in particular allow us to obtain the densities of
critical percolation clusters on the cylinder of the square lattice
of standard orientation  extensively studied before. We obtain exact densities of contractible
and non-contractible loops or equivalently the densities of critical
percolation clusters, which do not and do wrap around the cylinder
respectively. The solution uses the mapping of O(1) dense loop model
to the six-vertex model in the Razumov-Stroganov point, while the
effective tilt is introduced via the the inhomogeneous transfer matrix
proposed by Fujimoto. The further solution is based on the Bethe ansatz
and Fridkin-Stroganov-Zagier's solution of the Baxter's T-Q equation.
The results are represented in terms of the solution of two explicit
systems of linear algebraic  equations, which can be performed either
analytically for small circumferences of the cylinder or numerically
for larger ones. We present exact rational values of the densities
on the cylinders of small circumferences and several lattice orientations
and use the results of  high precision numerical calculations to study the finite-size
corrections to the densities, in particular their dependence on the
tilt of the lattice.
\end{abstract}
\maketitle

\section{Introduction}

Percolation is a classical problem used as a testing ground of the
theory of critical phenomena. It is formulated in terms of a graph, in which 
 bonds (for bond percolation) or sites (for site percolation)  are independently selected to be either open  or closed with fixed probabilities $\tt{p}$ and $(1-\tt{p})$  respectively.   The adjacent open  bonds or sites form connected clusters. The hallmark of the percolation at infinite graphs is a phase transition:  an infinite connected component arise at a critical point $\tt{p}=\tt{p}_c$. A significant effort has been paid to study this phase transition \cite{Grimmett, Kesten,Bollobas}. In particular, the density (the  mean  per site number) of connected clusters,   is one of simplest quantities that have been studied for a long time. This is the main subject of the present  paper.
 
A biggest progress in calculation of the cluster densities has been made   for the most analytically tractable  non-trivial  example of percolation, percolation on two-dimensional periodic planar lattices.  In early history the cluster densities  were investigated with the help of series expansions,
which in particular allowed the use of duality arguments to find critical
points for percolation in several lattices \cite{SykesEssam1963,SykesEssam1964}. A significant advance was achieved for the critical, $\tt{p}=\tt{p}_c$, bond percolation on the square lattice and   triangular lattices for them being exactly solvable
with the toolbox of the theory of integrable systems \cite{Baxter}.
In particular the exact values of the infinite plane limit of critical percolation cluster densities were obtained \cite{TL,BaxterTemperleyAshley1978}. 

These limits, however, being model dependent quantities  are not universal characteristics of the percolation phase transition. On the other hand, the scaling limit of the critical percolation in two dimensions is believed to enjoy the  conformal invariance that, in particular, was rigorously proved  for the site percolation on the triangular lattice \cite{Smirnov}. To see manifestation  of conformal invariance in the cluster densities
one can consider percolation in a restricted geometry, e.g. on the infinite lattice strip of a finite width.  
The conformal field theory (CFT) predicts  universal finite size corrections to the bulk values of the cluster densities, which depend  on the boundary conditions on the boundaries of the strip. They are given by the conformal anomalies \cite{BloteCardyNightingale1986}
that were found from using both the Bethe ansatz solution of related models \cite{ABB,HQB} or the Coulomb gas theory \cite{Nienhuis1987}.
 The CFT based leading finite size corrections to critical percolation cluster densities were conjectured and numerically checked in  \cite{ZiffFinchAdamchik,KlebanZiff}.

The mentioned results characterize the asymptotic behavior of cluster densities on a strip in the large strip width limit. On the other hand,  
the question about  exact values of the cluster  densities on strips of an arbitrary finite width  is still open even for  the simplest planar lattices like square and triangular lattice.
Their explicit expressions would allow one  not only to obtain the leading universal finite size corrections, but also to study the next terms of the asymptotic expansion in the 
inverse strip width, which  might also have a interesting CFT content.   

The dependence of the densities of percolation clusters on the open bond probability $\tt{p}$  was studied in \cite{ChangShrock2004,ChangShrock2021}
for lattice strips of finite width. Unfortunately,  the integrability toolbox can not be used for  general values $\tt{p}$. Therefore, only a few small values
of the strip width were treated, for which the problem is reduced to manipulations with the finite dimensional transfer matrix. No generalization
for arbitrary lattice sizes is known to be achievable in this way. 

\begin{figure}
\centerline{\includegraphics[width=0.3\textwidth]{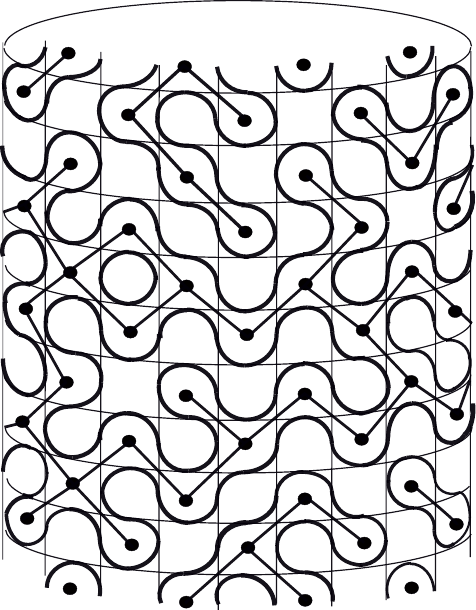}}
\caption{Loop configuration of O(1) DLM on the square lattice wrapped into a cylinder and corresponding percolation clusters on the forty five degree rotated   lattice. 	\label{loops_cylinder}}
\end{figure}

Here we are going to apply an integrability based approach  to study the critical, $\tt{p}=\tt{p}_c=1/2$, percolation cluster densities on the strip  
wrapped into a cylinder with boundary conditions specified below. To this end, we note that the percolation problem  can alternatively
be formulated in different languages, e.g.  as particular limits of Potts model or random cluster
model that in turn can be related to the  $O(n)$ dense loop model (DLM) 
 \cite{Baxter}. The O(n) DLM on the square lattice is defined as an ensemble of  weighted   non-crossing paths  passing  through every bond and making a ninety degree turn at  every site, so that  every closed loop brings the weight $n$. 
In particular, the critical bond percolation model on the square lattice can be mapped to the O(1) DLM on another square lattice that is a so called medial graph  of the original lattice.  Conversely, given a loop configuration on the square lattice, we can recover the percolation cluster configuration on the forty five degree rotated square lattice with vertices associated with half of  faces of the original lattice arranged in the checkerboard pattern.  In particular, the original  lattice (with loops) wrapped into a cylinder of even circumference corresponds to the rotated one (with percolation clusters)  also wrapped into a cylinder, see fig. \ref{loops_cylinder}. Thus, instead of studying the statistics of percolation clusters, one can equivalently study the statistics of loop configurations.
Specifically, as we explain later, the critical percolation cluster density is equal
to the density of loops on the medial lattice.   

The  solvability of the $O(1)$ DLM on the square lattice is due to its   connection 
with the Bethe ansatz solvable  six vertex model \cite{Baxter,TL,BaxterKellandWu}. In particular,  a special point of its  parameter space related to O(1)  DLM is the so
called Razumov-Stroganov  point, distinguished by a remarkable combinatorial structure of the ground state eigenvector of the six vertex transfer matrix or of  the related XXZ Hamiltonian  \cite{RS-01}. It is this point that is responsible for a possibility of obtaining  exact finite size formulas for the ground state observables of the model.
Indeed, connections between the percolation, the $O(1)$ DLM, the six vertex
model, the XXZ model, the fully packed loop model and alternating
sign matrices \cite{BGM,RS-04,RS-05,G} yielded plenty conjectures
and exact results  for finite lattices
with various boundary conditions \cite{FZ2005,FZ2005_2,Z2006,FZZ2006,FZ2007,RSZ2007,CS,GBNM,MNGB-1,GierJacobsenPonsaing,MitraNienhuis}.

However, the exact densities of loops in O(1) DLM and related critical percolation cluster densities had not been studied until recently.
In a recent letter \cite{Povolotsky2021} we considered the $O(1)$
dense loop model (DLM) on an infinite  the square
lattice cylinder of even circumference. Note that in the cylinder geometry one can distinguish between two types 
of loops, contractible and non-contractible ones, that can be given their own fugacities within the mapping 
to the six vertex model  \cite{AlcarazBrankovPriezzhevRittenbergRogozhnikov}.  Thus, the exact densities of contractible
and non-contractible loops, which also gave the densities of the critical
percolation clusters on the forty-five degree rotated lattice, were
obtained. We showed that these densities are given by explicit rational
functions of the circumference of the cylinder. At large circumference,
the leading and sub-leading orders of their asymptotic expansion reproduced
 the previous asymptotic results. 

The densities obtained, however, can not be directly compared to most
of the results on percolation available to date, e.g. those of \cite{ChangShrock2004,ChangShrock2021},
since the latter are obtained for the percolation on the cylinder with standard lattice orientation, 
which conversely would correspond to O(1) DLM on forty five degree rotated lattice.
In general, the infinite plane limit of the cluster densities on the
cylinder does not depend on the lattice orientation, and the form
of leading finite size correction to this limit is expected to be universal,
i.e. to depend on the lattice orientation  only via the length
rescaling. At the same time, the exact values of critical percolation
cluster densities and, in particular,
 the next to sub-leading  finite size corrections do depend on the lattice orientation. 
 This dependence is the subject of the present article.

Below we continue the studies started in \cite{Povolotsky2021} carrying out calculations of the densities of contractible and
non-contractible loops in $O(1)$ DLM on the cylinder obtained from
the square lattice rotated by an angle $\alpha$, such that $\tan\alpha=m/n$
is a rational number indexed by two co-prime integers $m\leq n$. Correspondingly
they yield the densities of critical percolation clusters on the cylinder of the square
lattice rotated by the angle $\alpha+\pi/4$ with respect to the standard
orientation. In particular, when $\alpha=\pi/4$, corresponding to
$(n,m)=(1,1)$, we obtain the cluster densities for the standard lattice
orientation, which can be compared with previous results.

The solution is based on the mapping of the $O(1)$ DLM to the six-vertex
model \cite{BaxterKellandWu}. To introduce the tilt into the lattice
we apply  the trick proposed in \cite{Fujimoto1994}, see also \cite{BY1995}, that consists in
use of the transfer matrix of inhomogeneous six vertex model, which
effectively replaces part of the vertices of the lattice by a pair
of non-interacting bonds. As a result, we arrive at the T-Q equation,
the solution of which at the Razumov-Stroganov point can be represented
in terms of the solution of an explicit linear system. Then, the densities
of clusters given by the derivatives of the largest eigenvalue of
the transfer matrix with respect to parameters is expressed in terms
of the Q-operator and the P-operator that solves a conjugated
T-P equation. The technique based on the interplay between P- and Q- operators 
comes back to Pronko and Stroganov \cite{PS1999} (see also \cite{BGJN2020} 
for the case with a twist). The method of exact solution of T-Q and T-P equations 
at the Razumov-Stroganov point and  of calculating derivatives of the largest eigenvalue with respect to the loop fugacities  
was developed by Fridkin, Stroganov, Zagier in \cite{FSZ-1,FSZ-2}. Our calculations follow the line of \cite{Povolotsky2021}
based on those two papers. 

Unlike \cite{Povolotsky2021}, here we were not able to express the
Q- and P-operators in terms of hypergeometric (or any other
special) functions. This is why the final result does not have an
explicit functional form. Rather it is given in terms of the solution
of the linear system. For a finite circumference of the cylinder it
can be solved using the Wolfram Mathematica, and yields explicitly
exact rational values of the densities for the circumferences as small
as a few tens and not too big values of $m$ and $n$. The approximate
decimal values of the densities calculated with arbitrarily high numerical
precision are available for much bigger sizes of the densities. We
use the decimal approximation to study the tilt dependence of the
finite size corrections to the densities.

The article us organized as follows. In section 2 we introduce the O(1) DLM, critical percolation model and  the six-vertex model on the tilted lattice, and explain  connections between them. Then, we formulate a problem of finding the specific free energy 
for these models, from which   the densities of loops and clusters can be obtained as its derivatives. In Section 3 we derive T-Q and T-P equations and show how the derivatives of the free energy can be obtained from Q- and P-operators. In Section 4 we give  solutions for   Q- and P-operators in terms of the solution of an explicit system of linear algebraic equations and use them to derive the final formulas for the densities. Section 5 contains final results on exact rational values of the densities for small lattices and asymptotic analysis of results of  numerical solution, which demonstrate  conformal invariance of the sub-leading finite size corrections and leads us to conjectures on the form of the non-universal next to sub-leading finite size corrections to the density.     

\section{O(1) DLM, percolation and six-vertex model on a tilted lattice. }

Let us fix two non-negative co-prime integers $m,n\in\mathbb{N}_{0}$, which
are not zero simultaneously, and  positive integer $l\in \mathbb{N}$, such that
\begin{equation}
	L=(m+n)l\in 2\mathbb{N} \label{eq: L}
\end{equation}
is an even positive integer. Though we can consider arbitrary $n,m$ and $l$ such that $L$ is even, transferring factors between the first two and the third one as well as  swapping $n$ and $m$ results in  equivalent situations.Therefore it is enough to limit our choice to co-prime $m \leq n$ and arbitrary  $l$.

Consider a strip of the square lattice $\mathcal{L}=(V,E)$ with vertex
set $V=\{1,\dots,ln\}\times\mathbb{Z}$ and edge (or bond) set $E=\{(v,v+\boldsymbol{e}_{x}),(v,v+\boldsymbol{e}_{y}))\}_{v\in V}$,
where $\boldsymbol{e}_{x}=(1,0)$ and $\boldsymbol{e}_{y}=(0,1)$
are lattice vectors, rolled into a cylinder with helical boundary
conditions (BC), i.e. we imply that $v\equiv v+nl\cdot\boldsymbol{e}_{x}-ml\cdot\boldsymbol{e}_{y}$
for any $v\in V$. The helical BC introduce a tilt with angle $\alpha$,
such that $\tan\alpha=m/n$. The $O(1)$ DLM on this lattice is formulated
as a measure on path configurations, in which a path passes through
every bond exactly once, and two paths meet at every site without
crossing each other. All the path configurations on any finite part
of the lattice have equal weights.

\begin{figure}[h]
\centering{}\includegraphics[width=0.15\textwidth]{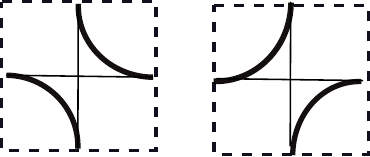}\caption{Two vertices of the O(1) loop models. Both vertices have unit weight.\label{fig:loop-vertices}}
\end{figure}

The path configurations can be constructed with local operations by
placing one of two vertices at every lattice site, in which two pairs
of paths at four incident bonds are connected pairwise in one of two
possible ways shown in fig. \ref{fig:loop-vertices}, both assigned
with the unit weight.

The choice of even $L$ ensures that under the uniform measure on
paths only finite closed loops present on the cylinder with probability
one, each loop having the weight $w=1$.

\begin{figure}
\centering{}\includegraphics[width=0.3\textwidth]{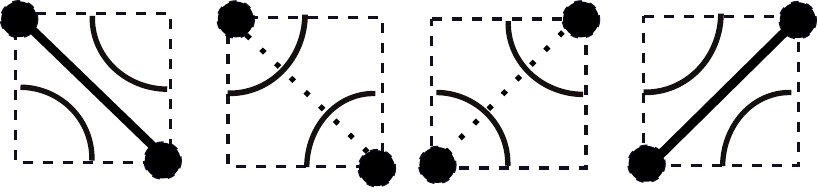}\caption{Correspondence between the vertices of O(1) DLM at $\mathcal{L}$
and open (solid) or closed (dashed) bonds on $\mathcal{L}'$. The
black dots are sites of $\mathcal{L}'.$ \label{fig:loop-perc}}
\end{figure}

\begin{figure}
\centering{}\includegraphics[width=0.4\textwidth]{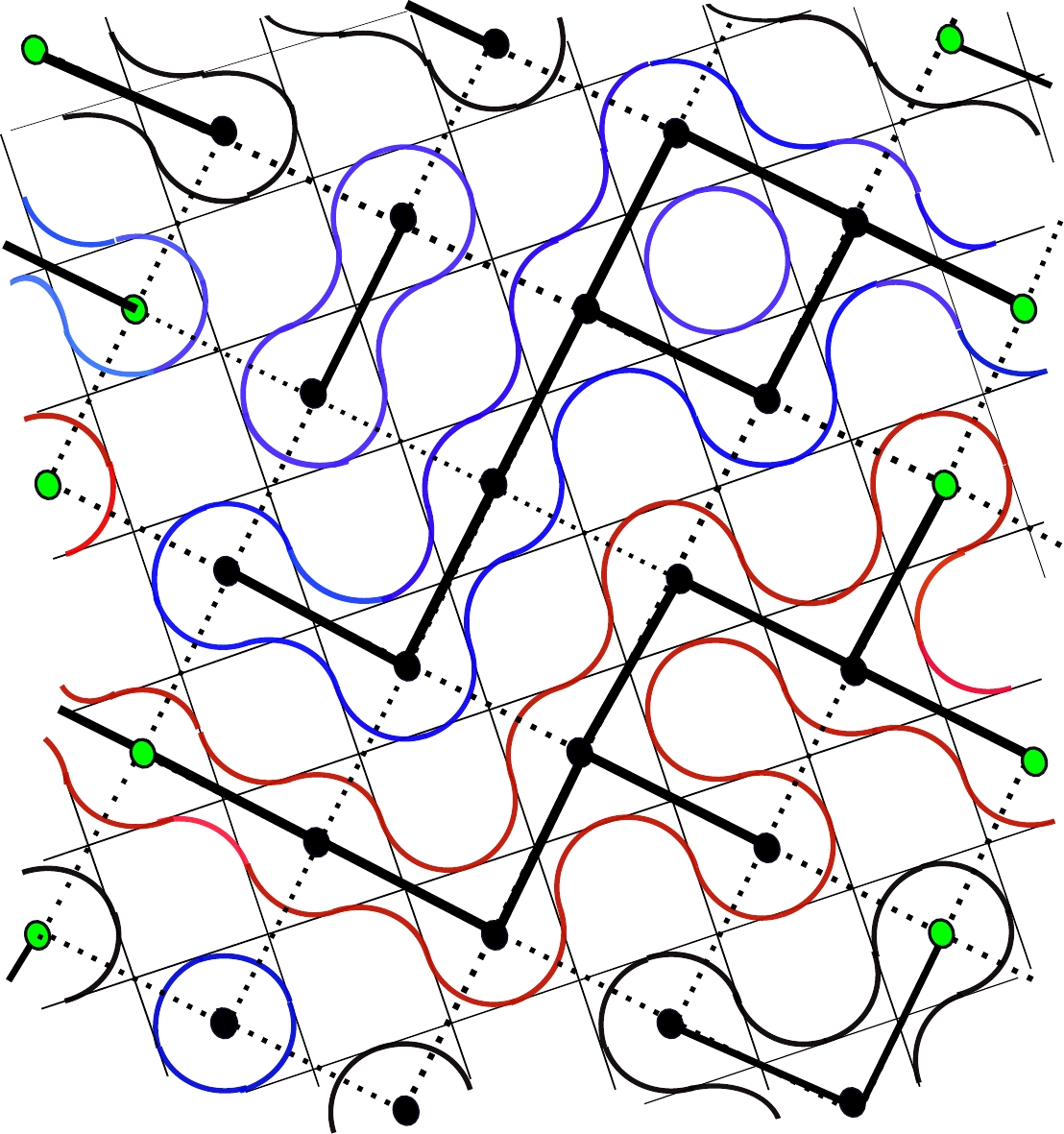}\caption{Loops on the part of infinite cylinder rolled out of a tilted lattice
with $(m,n)=(3,1)$ and associated percolation clusters on the lattice
rotated by $\pi/4$. Contractible and non-contractible loops are shown
in blue and red respectively. The green dots on the left and right
hand sides mark the faces (and the sites if the rotated lattice),
which are identified when the strip is rolled into the cylinder.\label{fig:lattice}}
\end{figure}

The loop configurations are in one to one correspondence with the
set of open and closed bonds on the lattice $\mathcal{L}'$, for which
the original lattice is the medial graph, i.e.  sites of $\mathcal{L}'$ are placed to
the center of every second face of $\mathcal{L}$  in a staggered
way and bonds are passing through the nearest sites of $\mathcal{L}$,
see fig. \ref{fig:lattice}. The helical boundary conditions for
$\mathcal{L}$ suggest that $\mathcal{L}'$ is also a cylinder rolled
out of the strip of the square lattice, but with the tilt $\alpha'=\alpha+\pi/4$.
In particular the choice $(n,m)=(1,1)$ suggests that $\alpha'=\pi/2$,
i.e. $\mathcal{L}'$ is the square lattice in the standard orientation.
Then, a bond of $\mathcal{L}'$ is open (closed), when it is between
(crosses) the loop arcs, see fig. \ref{fig:loop-perc}.  The probability of an open  bond is $1/2$ as well as of a closed one. This     is the critical point of the bond percolation
on the infinite square lattice. 

We use the notations $\nu_{\mathrm{c}}(l,m,n)$ and $\nu_{\mathrm{nc}}(l,m,n)$ for
the densities, i.e. average per site numbers, of contractible and
non-contractible loops, respectively. Similarly to \cite{Povolotsky2021}
they are also the densities of percolation clusters that do not and
do wrap around the cylinder $\mathcal{\mathcal{L}}'$ respectively.
This is obvious for   non-contractible loops since every  percolation cluster wrapped around the cylinder on $\mathcal{L}'$ is
surrounded by two non-contractible loops on $\mathcal{L}$, while $\mathcal{L}'$  contains twice less sites per unit
length of the cylinder than  $\mathcal{L}$.  
For contractible loops, we note that every contractible loop is either circumscribed
on a percolation cluster that does not wrap around the cylinder or
is inscribed into a hole inside a percolation cluster. The latter
loop can also be thought of as circumscribed on the dual percolation
cluster on the lattice dual to $\mathcal{L}'$. The critical point is self-dual.
This means that the average numbers of percolation clusters and of
dual percolation clusters are equal, and so are the average numbers
of the circumscribed and the inscribed loops.

\begin{figure}[h]
\centering{}\includegraphics[width=0.5\textwidth]{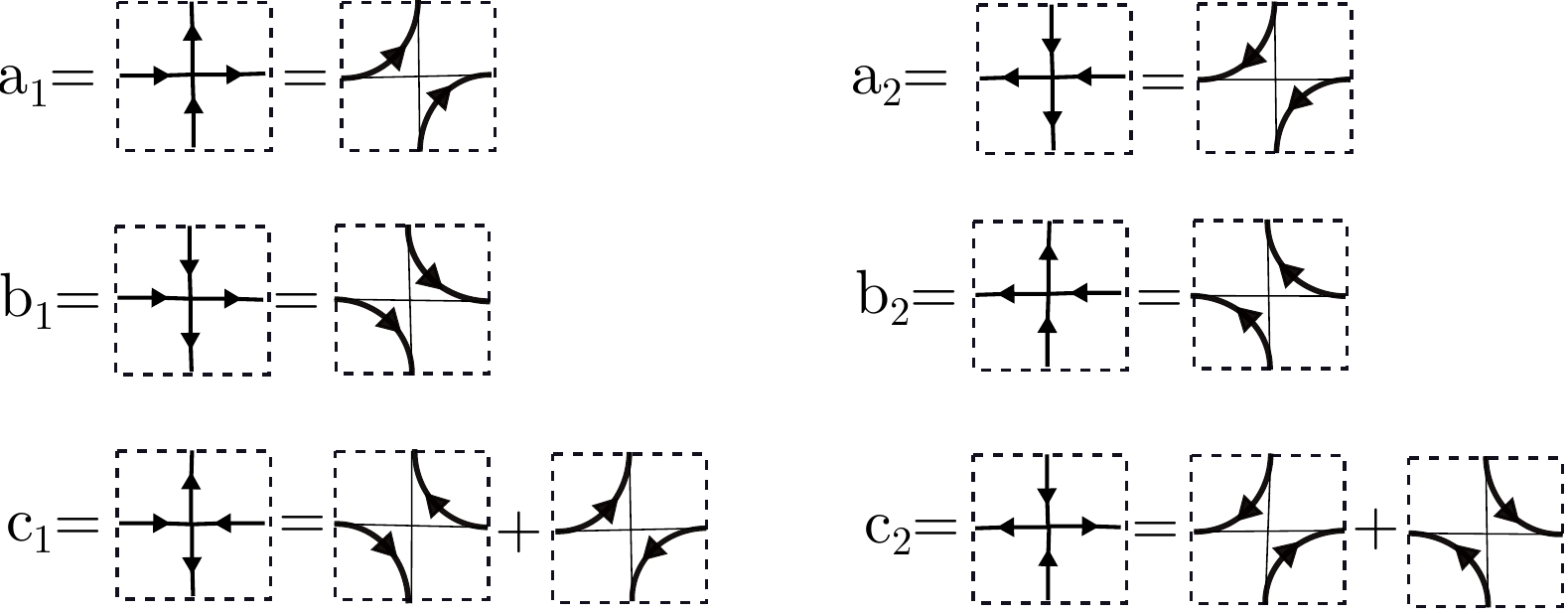}\caption{Correspondence between the six-vertex model and the directed loop
model. \label{fig:six vertex-loop}}
\end{figure}

To proceed with the analytic solution we exploit the relation of $O(1)$
DLM with the six-vertex model going first to the directed loop model
\cite{BaxterKellandWu} by giving either clockwise or counterclockwise
orientation to every loop. This makes the arcs within the vertices
in fig.~\ref{fig:loop-vertices} directed, which is indicated by an
arrow, fig.~\ref{fig:six vertex-loop}. Then, attaching arrows to bonds
incident to every site consistently with the directions of the arcs
and ignoring the connectivities we obtain six vertices of the six-vertex
model out of eight vertices of the directed loop model.

To define the model on the strip of even width $L$ of standard orientation square lattice rolled into a cylinder
 we assign the following weights to the vertices 
\begin{eqnarray}
a_{1} & = & \frac{u^{1/2}-u^{-1/2}}{q^{-1/2}-q^{1/2}}e^{\mathrm{i}\frac{\varphi}{L}},\quad a_{2}=\frac{u^{1/2}-u^{-1/2}}{q^{-1/2}-q^{1/2}}e^{-\mathrm{i}\frac{\varphi}{L}},\label{eq: a1,a2}\\
b_{1} & = & \frac{q^{-1}u^{-1/2}-qu^{1/2}}{q^{-1/2}-q^{1/2}}e^{\mathrm{i}\frac{\varphi}{L}},\quad b_{2}=\frac{q^{-1}u^{-1/2}-qu^{1/2}}{q^{-1/2}-q^{1/2}}e^{\mathrm{-i}\frac{\varphi}{L}},\label{eq: b1,b2}\\
c_{1} & = & q^{1/2}+q^{-1/2},\quad c_{2}=q^{1/2}+q^{-1/2}.\label{eq: c1,c2}
\end{eqnarray}
This is one of standard parametrizations of  the weights of  six-vertex model \cite{Z2009}. Then, at a special value of the spectral parameter $u=1/q$ the weight of clockwise (counterclockwise) quarter turn of a directed loop is given by  $q^{1/4}$ ($q^{-1/4}$) and the weight of the right (left) horizontal step is  $e^{\mathrm{i}\varphi/L}$ ($e^{-\mathrm{i}\varphi/L}$).  
As a consequence, the weights 
of contractible and non-contractible undirected loops are given
by 
\begin{equation}
w_{\mathrm c}=2_{q}=q+q^{-1}\quad\mathrm{and}\quad w_{\mathrm nc}=2 \cos\varphi\label{eq: loop weights}
\end{equation}
respectively becoming the unit weights, $w_{\mathrm c}=w_{\mathrm nc}=1,$ in the so called
stochastic point 
\begin{equation}
q=e^{\frac{\mathrm{i}\pi}{3}},\quad\varphi=\pi/3.\label{eq:stochastic}
\end{equation}
Let us define the $R$-matrix 
\begin{equation}
R(u)=\left(\begin{array}{cccc}
a_{1} & 0 & 0 & 0\\
0 & b_{1} & c_{1} & 0\\
0 & c_{2} & b_{2} & 0\\
0 & 0 & 0 & a_{2}
\end{array}\right)
\label{R-matrix}
\end{equation}
and its analogues acting in the tensor product of $L+1$ two-dimensional
spaces $R_{ij}(u)\in\mathrm{End}\left(\mathbb{C}^{2}\right)^{\otimes(L+1)}$,
which act as $R(u)$ in the pair of spaces $i$ and $j$ and identically
in the others. Such defined $R-$matrices satisfy the following Yang-Baxter
equation 
\begin{equation}
R_{12}\left(q^{-2}u/v\right)R_{13}\left(u\right)R_{23}\left(v\right)=R_{23}\left(v\right)R_{13}\left(u\right)R_{12}\left(q^{-2}u/v\right).\label{eq: Yang-Baxter}
\end{equation}

As was noted above, to construct the transfer matrix for the model
on the lattice in the standard orientation it would be enough to use
a particular specialization $R(1/q)$ of the $R$-matrix, which we
will refer to as normal vertices.

It was shown in \cite{Fujimoto1994,BY1995} that the use of auxiliary vertices
obtained from other specializations allows one to introduce an effective
tilt into the lattice. Note, that first solution of the problem of six-vertex model on the rotated lattice was given in  \cite{LitvinPriezzhev1990} using the
  method based on the random walk representation of the Bethe
ansatz. Here, we follow \cite{Fujimoto1996}, where the Reader can
consult about the details of the further construction.

\begin{figure}
\includegraphics[width=0.7\textwidth]{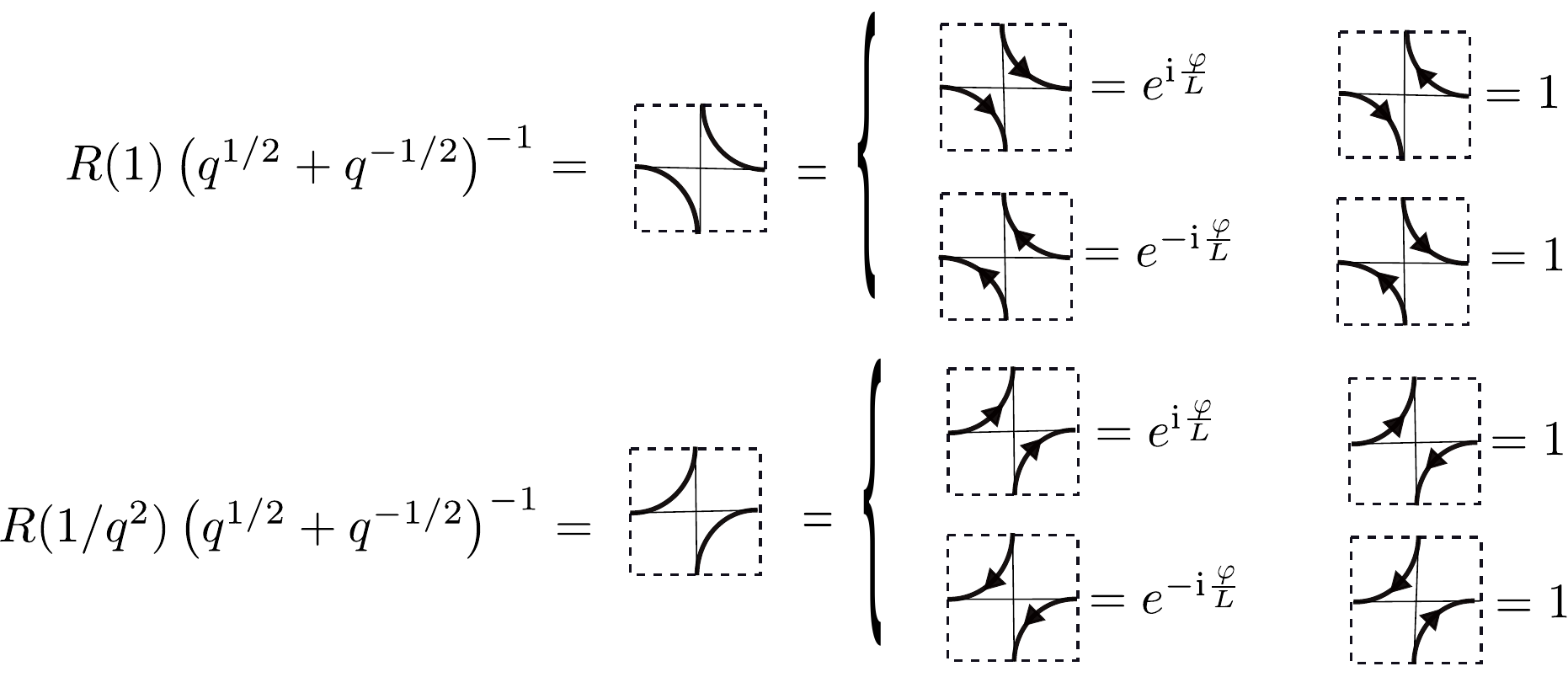}\caption{The auxiliary vertices.\label{fig:The-auxiliary-vertices.}}
\end{figure}

To define the tilted model, we first note that two other specializations
of the $R$-matrix 
\begin{equation}
R(1)=\left(q^{1/2}+q^{-1/2}\right)\left(\begin{array}{cccc}
0 & 0 & 0 & 0\\
0 & e^{\mathrm{i}\frac{\varphi}{L}} & 1 & 0\\
0 & 1 & e^{-\mathrm{i}\frac{\varphi}{L}} & 0\\
0 & 0 & 0 & 0
\end{array}\right)\label{eq: R(1)}
\end{equation}
and
\begin{equation}
R\left(q^{-2}\right)=\left(q^{1/2}+q^{-1/2}\right)\left(\begin{array}{cccc}
e^{\mathrm{i}\frac{\varphi}{L}} & 0 & 0 & 0\\
0 & 0 & 1 & 0\\
0 & 1 & 0 & 0\\
0 & 0 & 0 & e^{\mathrm{-i}\frac{\varphi}{L}}
\end{array}\right)\label{eq: R(q-2)}
\end{equation}
can be treated as vertices with pairs of arcs connecting either west
to south and north to east or west to north and east to south, see
fig. \ref{fig:The-auxiliary-vertices.}.

Up to the overall extra factor $(q^{1/2}+q^{-1/2})$ the matrix elements
corresponding to unit weight vertices with fixed connectivity are
also supplied with factors $e^{\pm\mathrm{i}\varphi/L},$ which account
for the left-right arrow propagation.

Given  $n,m,l$ and $L$ defined in (\ref{eq: L}), we introduce a one-parametric family of commuting transfer matrices
\begin{equation}
V(u)=\mathrm{Tr}_{0}\prod_{i=0}^{l-1}\left(\prod_{j=il+1}^{il+m}R_{0j}\left(\frac{u}{q}\right)\prod_{k=il+m+1}^{il+m+n}R_{0k}(u)\right),\label{eq: V(u)}
\end{equation}
which are operators acting in the tensor product of $L$ two-dimensional
``quantum'' spaces, $V(u)\in\mathrm{End}\left(\mathbb{C}^{2}\right)^{\otimes L}$,
with indices $1,\dots,L$, while the auxiliary space with index $0$ has
been traced out. The commutativity of the transfer matrices at different
values of the spectral parameter, 
\[
[V(u),V(v)]=0,
\]
can be proved using the Yang-Baxter equation (\ref{eq: Yang-Baxter}).

Thus, there are $l$ groups of $m$ auxiliary vertices of type (\ref{eq: R(q-2)})
alternating with $l$ groups of $n$ normal vertices in $V(1/q)$.
Likewise, $V(1)$ contains $l$ groups of $m$ normal vertices alternating
with $l$ groups of $n$ auxiliary vertices of type (\ref{eq: R(1)}).

Concatenating groups of $n$ rows of the lattice of type $V(1/q)$
alternatingly with groups of $m$ rows of type $V(1)$ and letting
the number of rows of the lattice to be $(m+n)l'$ with some $l'\in\mathbb{N}$
we define partition function 
\begin{equation}
Z_{n,m,l,l'}(q,\varphi)=\mathrm{Tr}_{1,\dots,L}\left[\left(V(1/q)\right)^{n}\left(V(1)\right)^{m}\right]^{l'}\left(q^{1/2}+q^{-1/2}\right)^{-2mnll'}.\label{eq: Z}
\end{equation}
which is nothing but the torus partition function of the loop model on the inhomogeneous lattice with  some  vertices  replaced by a pair of non-interacting arcs,  while contractible and non-contractible
loop weights are still as in (\ref{eq: loop weights}).\footnote{To be precise,  the weight of non-contractible loops  defined in (\ref{eq: loop weights})  
	applies only to loops  winding once in  the horizontal direction and  not winding around the other cycle of the torus.  These are the only non-contractible loops present on the infinite cylinder we finally aim at. Other non-contractible loops winding around both cycles of the torus, which  assign   weight $2\cos(n \varphi)$ to a loop winding $n$ times around the horizontal cycle, also contribute to the torus partition function. These loops, however,   turn to infinite loops (defects), when the second cycle is sent to infinity. The loop configurations with such defects have zero measure in O(1) DLM on an infinite cylinder and, in particular, do not affect the free energy obtained form the subsequent  $l'\to\infty$ limit.} 
One can see in fig. 7  that the new lattice obtained by introducing the  split  vertices  is equivalent to the tilted lattice, while the periodic boundary conditions turn into the helical boundary conditions described in the beginning of the  section. 
\begin{figure}
	\centerline{	\includegraphics[width=0.8\textwidth]{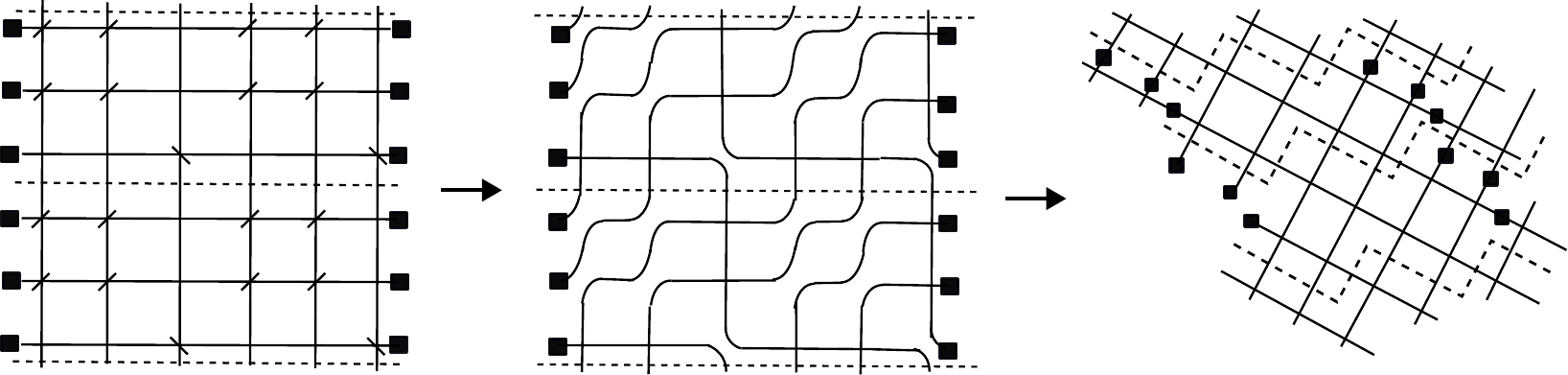}}
	\caption{Transformation of the lattice by introducing the auxiliary vertices  split  into two non-interacting arcs.  The example shown corresponds to $(n,m)=(2,1) $ and $l=2$. The  cuts in  split vertices corresponding to $R(1/q^2)$ and $R(1)$ are shown by  $\slash$  and  $\backslash$  respectively on the leftmost picture, while the usual vertices correspond to $R(1/q)$. The split vertices are replaced by pairs of arcs in picture in the center. The stretch of the central picture converts it  to the piece of the tilted square lattice on the right. The left and right  black squares  on the same horizontal level are identified under imposing periodic boundary due to the    trace in (\ref{eq: V(u)}). 
	 The strip between two dashed horizontal lines  on the left and central pictures  becoming the zig-zag-like strip in the right one corresponds to  the transfer matrix $(V(1/q))^n(V(1))^m$ that acts between two spaces depicted by the dashed lines. The strips are  concatenated repeatedly $l'$ times in the vertical direction and the first  space is identified with the last space   within the partition function (\ref{eq: Z}).
		 \label{tilted}  }
\end{figure}
Here, the second
factor compensates for the extra weight coming from $2mnll'$ auxiliary
vertices. Now we define the per site free energy of the model on infinite cylinder,
obtained by letting $l'\to\infty,$
\begin{equation}
f_{n,m,l}(w_{\mathrm c},w_{\mathrm nc})=\lim_{l'\to\infty}\frac{\log Z_{n,m,l,l'}(q,\varphi)}{\left(m^{2}+n^{2}\right)ll'},\label{eq: free energy}
\end{equation}
that is a function of the loop weights defined in (\ref{eq: loop weights}) as well as of the parameters $n,m$ and $l$. Here the denominator $\left(m^{2}+n^{2}\right)ll'$ is the number
of normal vertices out of the total number $(m+n)^{2}ll'$ of the
vertices. In particular the contractible and non-contractible loop
densities are given by

\begin{eqnarray}
\nu_{\mathrm{c}} (l,m,n) &= & \left.w_{\mathrm{c}}\frac{\partial}{\partial w_{\mathrm c}}\right|_{w_{\mathrm c},w_{\mathrm nc}=1}f_{n,m,l}(w_{\mathrm c},w_{\mathrm nc}), \label{eq:f-deriv_c} \\
\nu_{\mathrm{nc}}(l,m,n)&=&\left.w_{\mathrm{nc}}\frac{\partial}{\partial w_{\mathrm nc}}\right|_{w_{\mathrm c},w_{\mathrm nc}=1}f_{n,m,l}(w_{\mathrm c},w_{\mathrm nc}).\label{eq:f-deriv_nc}
\end{eqnarray}

To evaluate the free energy we note that the  limit in (\ref{eq: free energy}) is governed by the largest eigenvalue of the transfer matrix (\ref{eq: V(u)}). To identify the largest eigenvalue we first note that the space $\mathcal{H}=\left(\mathbb{C}^{2}\right)^{\otimes L}$,  where the transfer matrix $V(u)$ acts,  is a span of the basis consisting  of $2^L$ vectors of the form $\otimes_{i=1}^L e^{(k_i)}_{i}$, where   
a factor $e^{(k_i)}_{i}$  with the superscript $k_i\in\{0,1\}$ is one of the vectors $e^{(0)}=(0,1)^T$ or $e^{(1)}=(1,0)^T$ and the running subscript $i$ indexes the spaces within the tensor product.	Such  basis vectors represent  possible states of   $L$ vertical bonds on the same horizontal level of the lattice with up or down arrow  corresponding to $e^{(0)}$ and $e^{(1)}$ respectively.
Then, the whole space $\mathcal{H} $   is decomposed  into $L+1$ invariant sub-spaces $\mathcal{H}=\oplus_{p=0}^{L}\mathcal{H}_p$, where each $\mathcal{H}_p$ stable under the action of $V(u)$ is a span of the basis  vectors with fixed number $p=0,\dots,L$  of up arrows. Then, the question is  in which subspace the eigenstate corresponding to the largest eigenvalue lives  and whether it is unique.  

The first answer to this question applied to variants of the symmetric six vertex model  was given in  Lieb's seminal solution of the ice, KDP and F models  \cite{Lieb1,Lieb2,Lieb3,Lieb4}  based on the results of Yang and Yang \cite{YY1,YY2} for the related XXZ chain. In particular, Lieb  established that in the     disordered phase,  $-1<\Delta<1$, of the symmetric six-vertex model on the cylinder of circumference $L$  the  dominant  eigenstate  belongs to the invariant subspace $\mathcal{H}_p$ with  $p=p_{\max}$    asymptotically equal to  $L/2$,   i.e. $p_{\max}/L\to 1/2$ as $L\to \infty$.    Furthermore, in that case     the transfer matrix  restricted to a subspace $\mathcal{H}_p$ with any $p$  is  irreducible and aperiodic  with real non-negative coefficients, i.e. satisfying  conditions of the Perron-Frobenius theorem. Hence, the maximal eigenstate in each invariant  subspace  is the  Perron-Frobenius eigenvector, which is non-degenerate with strictly positive components.  This is   true in particular for the maximal  over all the invariant sub-spaces eigenvalue. To our knowledge, the  statement about $p_{\max}$ is still proved rigorously only asymptotically, see e.g. \cite{D-CKKMT}, though a commonly believed conjecture is that a stronger statement can be made   that holds for finite $L$. In particular,   the non-degenerate largest eigenvalue is expected to be found in the subspace with  $p=p_{\max}=L/2$ for even finite $L$. 

In our notations, the symmetric six vertex model in the disordered phase with $\Delta=-(q+q^{-1})/2$ corresponds to the weights    (\ref{eq: a1,a2}-\ref{eq: c1,c2}) with $\varphi=0$ and $|q|=|u|=1$ under constraints  $0<\arg q< \pi$ and $0<2 \arg q+\arg u<2 \pi$, which  ensure non-negativity of Boltzmann weights.  Thus,  it would  be natural to expect that the properties mentioned are preserved by continuity at least in some vicinity of $\varphi=0$, though at nonzero $\varphi$ the  transfer matrix is not  real valued anymore. 
In fact,  the situation turns out even better at the stochastic point (\ref{eq:stochastic}) exactly due to its connection with O(1) DLM
that can effectively be formulated as a  Markov chain, of which  the dominant eigenstate is simply a stationary state. 
The rough idea is to consider the transfer matrix of the homogeneous six vertex model in a different basis that was constructed in  \cite{MNGB-1,FZ2005,FZZ2006}. In this basis the six vertex transfer matrix turns into a real valued symmetric O(1) DLM transfer matrix  that adds a row  of vertices from fig.~\ref{fig:loop-vertices} on top of the semi-infinite cylinder. These basis vectors  record how the vertical terminal bonds are connected to one another by half-loops  within the semi-infinite cylinder. Thus, they are indexed  by non-crossing pairings the of $L$  points on the rim of punctured annulus and the basis spans the subspace $\mathcal{H}_{p}$ with $p=L/2$ of the arrow basis of the six-vertex model, while the other sub-spaces are related with the loop configurations containing  defect lines and  are suppressed in the infinite system.  In this setting the dominant eigenstate is the Perron-Frobenius eigenvector with the eigenvalue that is non-degenerate and thus analytic in its parameters at least in a vicinity of the stochastic point. In fact,  as far as   
spectral parameter $u$ varies in a range of values of $u$, which preserves 
the conditions of the Perron-Frobenius theorem, the largest eigenvalue remains non-crossing and the dominant eigenstate does not change being independent of $u$. See  \cite{MNGB-1,FZ2005,FZZ2006} for further details.

The same construction is readily applied  to our tilted transfer-matrix.  Specifically, the largest eigenvalue of $V(u)$ is a Perron-Frobenius eigenvalue in the range $-\pi/3 \leq \arg u\leq 0$, i.e. it is non-degenerate and thus analytic in $q$ and $\varphi$ in a vicinity of stochastic point (\ref{eq:stochastic})  and  the largest eigenstate belongs to the subspace  $\mathcal{H}_{L/2}$.  Taking  into account (\ref{eq: V(u)},\ref{eq: Z},\ref{eq: free energy}),
 the fact that the commuting transfer matrices are diagonalizable or at least can be brought to the Jordan form in the same  basis and also share the same non-degenerate dominant eigenvector, we express the free energy in terms of the largest eigenvalue $\Lambda_{\max}(u)$ of the transfer matrix $V(u)$ 
\begin{equation}
f_{n,m,l}(w_{\mathrm c},w_{\mathrm nc})=\frac{n\log\Lambda_{\max}(1/q)+m\log\Lambda_{\max}(1)}{\left(m^{2}+n^{2}\right)l}-\frac{2nm\log\left(q^{1/2}+q^{-1/2}\right)}{m^{2}+n^{2}}.\label{eq: f_n,m,l(w,v)}
\end{equation}
The next step of our program is to evaluate $\Lambda_{\max}(u)$ and its derivatives  with respect to $q$ and $\varphi$. 

\section{Bethe ansatz, T-Q, T-P and Q-P relations.}

The standard technique  of solution of the  eigenvalue problem for the transfer matrix of the six-vertex model is the  algebraic Bethe ansatz.
 The Bethe ansatz provides a recipe of construction  of   eigenstates   starting from the vacuum state with only down
arrows. Omitting standard though technical details that can be found e.g. in \cite{Faddeev1996},  we arrive at the following expression of the eigenvalue under assumption that corresponding eigenstate belongs to the subspace $\mathcal{H}_p\subset \mathcal{H}$ with $p=0,\dots,L$ up arrows 
\begin{eqnarray}
\Lambda(u) & =\frac{e^{\mathrm{i}\varphi}}{(-q)^{p}}\left(\frac{\left(u/q\right)^{1/2}-\left(q/u\right)^{1/2}}{q^{-1/2}-q^{1/2}}\right)^{ml}\left(\frac{u^{1/2}-u^{-1/2}}{q^{-1/2}-q^{1/2}}\right)^{nl}\prod_{k=1}^{p}\frac{q^{2}u-u_{k}}{u-u_{k}}\label{eq:Lambda(u)}\\
 & +\frac{e^{\mathrm{-i}\varphi}}{(-q)^{p}}\left(\frac{\left(uq\right)^{-1/2}-\left(uq\right)^{1/2}}{q^{-1/2}-q^{1/2}}\right)^{ml}\left(\frac{\left(u^{1/2}q\right)^{-1}-\left(u^{1/2}q\right)}{q^{-1/2}-q^{1/2}}\right)^{nl}\prod_{k=1}^{p}\frac{u-q^{2}u_{k}}{u-u_{k}},\nonumber 
\end{eqnarray}
where $u_{1},\dots,u_{p}$ are the roots of the system of Bethe ansatz
equations (BAE) 
\begin{equation}
e^{2\mathrm{i}\varphi}\left(\frac{u_{i}-q}{1-qu_{i}}\right)^{ml}\left(\frac{u_{i}-1}{q^{-1}-qu_{i}}\right)^{nl}=\left(-1\right)^{p-1}\prod_{k\neq i;k=1}^{p}\frac{u_{i}-q^{2}u_{k}}{u_{k}-q^{2}u_{i}}.\label{eq: BAE}
\end{equation}
The system (\ref{eq: BAE}) is exactly the condition of
$\Lambda(u)u^{(n+m)l/2}$ being polynomial in $u$, which follows
from the structure of the weights (\ref{eq: a1,a2}-\ref{eq: c1,c2}).

Let us rephrase it in a more convenient way.  First we note that the largest eigenvalue we are interested in  is known to belong to the sector with 
\[
p=\frac{L}{2}=\frac{l(n+m)}{2}.
\]
Thus, from now on we will assume this value of $p.$ 

For convenience we
introduce the following function 
\begin{equation}
T(u)=\Lambda(u)u^{p}q^{-p-nl/2}(-1)^{p}(1-q)^{L},\label{eq: T vs Lambda}
\end{equation}
which is expected to be a polynomial in $u$ of degree at most $L$.
We also introduce Q-polynomial
\[
Q(u)=\prod_{i=1}^{p}(u-u_{k}),
\]
of degree $p$ with roots being the Bethe roots from the solution
of (\ref{eq: BAE}) corresponding to the largest eigenvalue. Then
(\ref{eq:Lambda(u)}) can be rewritten in the form of T-Q equation
\begin{equation}
T(u)Q(u)=e^{\mathrm{i}\varphi}\Phi\left(uq^{-1}\right)Q(uq^{2})+e^{-\mathrm{i}\varphi}\Phi\left(uq\right)Q(uq^{-2}),\label{eq: TQ-equation}
\end{equation}
where 
\begin{equation}
\Phi(u)=(u-1)^{ml}(u-1/q)^{nl}.\label{eq: Phi}
\end{equation}
Solving this functional equation for polynomials $Q(u)$ and $T(u)$
of degrees $p$ and $L$ respectively is equivalent to solving BAE.

Note that an alternative route we could take, when going through the
Bethe ansatz procedure, is to use the state with all up arrows as
the vacuum. This is equivalent to solving the original six-vertex
model with the weights $a_{1},b_{1},c_{1}$ and $a_{2},b_{2},c_{2}$
from (\ref{eq: a1,a2}-\ref{eq: c1,c2}) exchanged respectively, which
is the same as the change $\varphi\leftrightarrow-\varphi$, while
the eigenvalue $\Lambda(z)$ (and hence $T(z)$) is the
same. This lead us to a functional relation 

\begin{equation}
T(u)P(u)=e^{\mathrm{-i}\varphi}\Phi\left(uq^{-1}\right)P(uq^{2})+e^{\mathrm{i}\varphi}\Phi\left(uq\right)P(uq^{-2})\label{eq:T-P}
\end{equation}
between yet another polynomial $P(u)$, which also has the degree
$p$ in the sector with $p=L/2$ up and down arrows, and $T(u)$.
Multiplying (\ref{eq: TQ-equation}) and (\ref{eq:T-P}) by $P(u)$
and $Q(u)$, equating the results, comparing zeroes of both sides
and  taking into account that $\Phi(u)\asymp u^{L}$ as $u\to\infty$, we
find that polynomials $P(u)$ and $Q(u)$ satisfy quantum Wronskian
relation 

\begin{equation}
\Phi(u)=\frac{e^{\mathrm{i}\varphi}Q(qu)P(q^{-1}u)-e^{-\mathrm{i}\varphi}Q(q^{-1}u)P(qu)}{e^{\mathrm{i}\varphi}-e^{-\mathrm{i}\varphi}}.\label{eq:phi-PQ}
\end{equation}
Substituting this back into any of (\ref{eq: TQ-equation},\ref{eq:T-P})
we find 
\begin{equation}
T(u)=\frac{e^{\mathrm{2i}\varphi}Q(q^{2}u)P(q^{-2}u)-e^{-2\mathrm{i}\varphi}Q(q^{-2}u)P(q^{2}u)}{e^{\mathrm{i}\varphi}-e^{-\mathrm{i}\varphi}}.\label{eq:T-PQ}
\end{equation}

To proceed with our final goal, calculation of the densities with
(\ref{eq:f-deriv_c},\ref{eq:f-deriv_nc}), we collect (\ref{eq: loop weights},\ref{eq:stochastic},\ref{eq: f_n,m,l(w,v)},\ref{eq: T vs Lambda})
to find 

\begin{eqnarray}
 & \nu_{\mathrm{c}}(l,m,n)=\frac{1}{\left(1-q^{-2}\right)l(m^{2}+n^{2})}\left[\frac{m}{T(1)}\left.\frac{dT(u)}{dq}\right|_{u=1}+\frac{n}{T(1/q)}\left.\frac{dT(u)}{dq}\right|_{u=1/q}\right]\label{eq: nu_c}\\
 & +\frac{1}{m^{2}+n^{2}}\left(\frac{1}{1-q^{-2}}\left(\frac{(m+n)(mq+m-nq+3n)}{2(1-q)q}-\frac{m}{q(q+1)}-\frac{n}{2q}\right)-\frac{mn}{3}\right) & ,\nonumber 
\end{eqnarray}

where $q=e^{\mathrm{i}\pi/3}$ should be substituted in the end, and
\begin{eqnarray}
\nu_{\mathrm{nc}}(l,m,n) & \frac{-1}{\sqrt{3}l(m^{2}+n^{2})}\left[\frac{m}{T(1)}\left.\frac{dT(1)}{d\varphi}\right|_{\varphi=\pi/3}+\frac{n}{T(1/q)}\left.\frac{dT(1/q)}{d\varphi}\right|_{\varphi=\pi/3}\right] & .\label{eq: nu_nc}
\end{eqnarray}

As shown in \cite{FSZ-1,FSZ-2}, the derivatives of $T(u)$ at the
stochastic point can be calculated, once those of $Q(u)$ and $P(u)$
are known. Indeed, setting $\varphi=\pi/3$ and differentiating (\ref{eq:T-PQ})
in $q$ at $q=e^{\mathrm{i}\pi/3}$ we obtain 
\[
\left.\frac{dT(u)}{dq}\right|_{q=e^{\mathrm{i}\pi/3},\varphi=\pi/3}=2A(u)+B(u)
\]
 where 

\begin{eqnarray}
A(u)&= & \frac{u}{q-q^{-1}}\Biggl(q^{\mathrm{2}}\left(qQ'(q^{2}u)P(q^{-2}u)-q^{-3}Q'(q^{2}u)P(q^{-2}u)\right)\label{eq: A(u)}\\
 & &-q^{-2}\left(qQ(q^{-2}u)P'(q^{2}u)-q^{-3}Q'(q^{-2}u)P(q^{2}u)\right)\Biggr)_{q=e^{\mathrm{i}\pi/3}} ,\nonumber 
\end{eqnarray}
is expressed in terms of the derivatives of $Q(u)$ and $P(u)$, while
$B(u)$ comes from the differentiation of the implicit dependence
of the Bethe roots on $q$. At the same time differentiating (\ref{eq:phi-PQ})
yields
\begin{equation}
\left.\frac{d\Phi(u)}{dq}\right|_{q=e^{\mathrm{i}\pi/3},\varphi=\pi/3}=-A(-u)+B(-u).\label{eq:d phi/dq}
\end{equation}
Eliminating $B(u)$ between (\ref{eq: A(u)}) and (\ref{eq:d phi/dq})
we obtain 
\begin{equation}
\left.\frac{dT(u)}{dq}\right|_{q=e^{\mathrm{i}\pi/3},\varphi=\pi/3}=3A(u)+\left.\frac{d\Phi(-u)}{dq}\right|_{q=e^{\mathrm{i}\pi/3}}.\label{eq: dT/dq}
\end{equation}
Similarly, for the derivative in $\varphi$ we obtain 
\begin{equation}
\left.\frac{dT(u)}{d\varphi}\right|_{q=e^{\mathrm{i}\pi/3},\varphi=\pi/3}=3C(u)+\frac{\Phi(-u)-T(u)}{\sqrt{3}},\label{eq: dT/dphi}
\end{equation}
where 
\begin{equation}
C(u)=2\mathrm{i}\frac{q^{2}Q\left(q^{2}u\right)P\left(q^{-2}u\right)+q^{-2}Q\left(q^{-2}u\right)P\left(q^{2}u\right)}{q-q^{-1}}\Biggr|_{q=e^{\mathrm{i}\pi/3},\varphi=\pi/3}.\label{eq: C(u)}
\end{equation}

\section{FSZ solution and densities of loops and clusters}

To apply the formulas obtained we need to find the solution of the
T-Q and P-Q equations at the stochastic point. To this end, we use
the observation made by Fridkin, Stroganov and Zagier (FSZ) about
the structure of this solution \cite{FSZ-1,FSZ-2} . Specifically,
for $q^{2}$ being the cube root of unity, the multiplicative shift
of the argument of T-Q equation by $q^{2}$ returns back in three
steps leading to a linear homogeneous system of tree equations. Indeed,
let us introduce notations 
\[
Q_{k}=Q\left(uq^{2k}\right),T_{k}=T\left(uq^{2k}\right),\Phi_{k}=\Phi(q^{k}u).
\]
 Then, the system obtained will read 
\begin{equation}
\boldsymbol{M}\boldsymbol{Q}=0,\label{eq: MQ system}
\end{equation}
where $\boldsymbol{Q}=(Q_{0},Q_{1},Q_{2})^{T}$ and 
\begin{eqnarray}
\boldsymbol{M} & = & \left(\begin{array}{ccc}
-T_{0} & e^{\mathrm{i}\varphi}\Phi_{-1} & e^{-\mathrm{i}\varphi}\Phi_{1}\\
e^{\mathrm{-i}\varphi}\Phi_{3} & -T_{1} & e^{\mathrm{i}\varphi}\Phi_{1}\\
e^{\mathrm{i}\varphi}\Phi_{3} & e^{-\mathrm{i}\varphi}\Phi_{5} & -T_{2}
\end{array}\right)\label{eq: M}
\end{eqnarray}
For the homogeneous system (\ref{eq: MQ system}) to be solvable the
rank of matrix $\boldsymbol{M}$ has to be at most two. The discovery
of FSZ was that for the ground state of the six-vertex model at the
stochastic point it equals to one, which in turn requires relation
\[
T_{k}=\Phi_{3+2k}
\]
to hold. It follows that at the stochastic point 
\begin{equation}
T(u)=\Phi(-u)\label{eq:T(u)=00003DPhi(-u)}
\end{equation}
where $\Phi(u)$ is as defined in (\ref{eq: Phi}). Expectedly, the
same argument applied to T-P equation, yields the same result. After
substitution to (\ref{eq: f_n,m,l(w,v)}) the latter result yields
the value of the free energy at the stochastic point

\[
f_{n,m,l}(1,1)=\log2,
\]
that is a consequence of the fact that there is a choice from two
weight one arc arrangements at every normal vertex independently of
the others. However, this is not yet our final goal, which requires
the derivative of the free energy with respect to the arguments. 

Unfortunately the further interpolation-like argument of FSZ leading
them to explicit formulas for the Q- and P-polynomials fails
in our case because of more complex explicit form of $\Phi(u)$. However,
one can find both Q- and P-polynomials solving explicitly the
linear systems for their coefficients. Specifically, following \cite{Motegi2013}
we define coefficients $q_{0},\dots,q_{p}$ and $p_{0},\dots,p_{p}$
as
\begin{eqnarray}
Q(u) & = & \sum_{k=0}^{p}q_{k}u^{k},\quad P(u)=\sum_{k=0}^{p}p_{k}u^{k},\label{eq: Q,P-coef}
\end{eqnarray}
 where by definition $q_{p}=p_{p}=1.$ It follows from (\ref{eq: TQ-equation})
and (\ref{eq:T-P}) at the stochastic point that each of the two sets
of $p$ unknown coefficients satisfy the following systems of $p$
equations
\begin{eqnarray}
\sum_{r=\max(0,3k+1-p)}^{\min(2p,3k+1)} & \phi_{r}q_{3k+1-r}(-1)^{r} & =0,\quad k=0,\dots,p-1,\label{eq: q-coef system}\\
\sum_{r=\max(0,3k+2-p)}^{\min(2p,3k+2)} & \phi_{r}p_{3k+2-r}(-1)^{r} & =0,\quad k=0,\dots,p-1,\label{eq: p-coef system}
\end{eqnarray}
where $\phi_{k}$ are coefficients of 
\[
\Phi(u)=\sum_{k=0}^{2p}\phi_{k}u^{k},
\]
given by 
\[
\phi_{k}=\left(-q\right)^{k-nl}\sum_{s=\max(0,k-nl)}^{\min(k,ml)}\left(\begin{array}{c}
nl\\
k-s
\end{array}\right)\left(\begin{array}{c}
ml\\
s
\end{array}\right)q^{-s}.
\]
Having solved this system for any finite $p$ we construct $Q(u)$
and $P(u)$, which can be substituted to formulas 
\begin{eqnarray*}
\!\!\!\!\!\!\!\!\!\!\!\!\!\!\!\!\!\!\!\!\!\!\!\!\!\!\!\!\!\!\nu_{\mathrm{c}}(l,m,n) & = & \frac{1}{m^{2}+n^{2}}\left(\frac{\mathrm{i}\sqrt{3}q^{ln+1}}{l}\left(\frac{m}{2^{lm}(q+1)^{ln}}A(1)+\frac{nq^{lm}}{2^{ln}(q+1)^{lm}}A\left(1/q\right)\right)+\frac{(m+n)^{2}}{2}\right)
\end{eqnarray*}

\[
\!\!\!\!\!\!\!\!\!\!\!\!\!\!\!\!\!\!\!\!\!\!\!\!\!\!\!\!\!\!\!\!\nu_{\mathrm{nc}}(l,m,n)=-\frac{\mathrm{i}\sqrt{3}}{l\left(m^{2}+n^{2}\right)}\left(\frac{m}{2^{lm}\left(1/q+1\right)^{ln}}C(1)+\frac{n}{\left(1/q+1\right)^{lm}\left(2/q\right)^{ln}}C\left(1/q\right)\right),
\]
obtained from (\ref{eq: nu_c}-\ref{eq: C(u)}) and (\ref{eq:T(u)=00003DPhi(-u)}). 

\section{Results and discussion}

Constructed formulas allowed us to reduce calculations of the densities
to solution of two linear systems. This can be done either analytically
or numerically with computer algebra systems. We performed the analytic
evaluation of a few exact densities. The values of $\nu_{\mathrm{c}}(l,0,1)$
and $\nu_{\mathrm{nc}}(l,0,1)$ confirm the general formula obtained in \cite{Povolotsky2021}
for the angle $\alpha=0$. In Table \ref{tab:Exact-densities} we
show the exact rational values of the densities of contractible and
non-contractible loops $\nu_{\mathrm{c}}(l,1,1)$ and $\nu_{\mathrm{nc}}(l,1,1)$ respectively
and also of the full density 
\[
\nu(l,m,n)=\nu_{\mathrm{c}}(l,m,n)+\nu_{\mathrm{nc}}(l,m,n)
\]
 of loops evaluated at $m=1$ and $n=1$ for $l=1,  \dots,10$. 

\begin{table}
\begin{tabular}{|l|l|l|l|}
\hline 
$l$ & $\nu_{\mathrm{c}}(l,1,1)$ & $\nu_{\mathrm{nc}}(l,1,1)$ & $\nu(l,1,1)$\tabularnewline
\hline 
\hline 
\noalign{\vskip\doublerulesep}
1 & $\frac{1}{6}$ & $\frac{1}{3}$ & $\frac{1}{2}$\tabularnewline[\doublerulesep]
\noalign{\vskip\doublerulesep}
\hline 
\noalign{\vskip\doublerulesep}
2 & $\frac{13}{110}$ & $\frac{9}{110}$ & $\frac{1}{5}$\tabularnewline[\doublerulesep]
\noalign{\vskip\doublerulesep}
\hline 
\hline 
\noalign{\vskip\doublerulesep}
3 & $\frac{1423}{13338}$ & $\frac{229}{6669}$ & $\frac{11}{78}$\tabularnewline[\doublerulesep]
\noalign{\vskip\doublerulesep}
\hline
\noalign{\vskip\doublerulesep}
4 & $\frac{1113499}{10834754}$ & $\frac{405855}{21669508}$ & $\frac{677}{5572}$\tabularnewline[\doublerulesep]
\noalign{\vskip\doublerulesep}
\hline 
\noalign{\vskip\doublerulesep}
5 & $\frac{5979030577}{59179172262}$ & $\frac{1747404017}{147947930655}$ & $\frac{85013}{753370}$\tabularnewline[\doublerulesep]
\noalign{\vskip\doublerulesep}
\hline 
\noalign{\vskip\doublerulesep}
6 & $\frac{217910906936461}{2176660978677230}$ & $\frac{17718816661443}{2176660978677230}$ & $\frac{1996408}{18442085}$\tabularnewline[\doublerulesep]
\noalign{\vskip\doublerulesep}
\hline 
\noalign{\vskip\doublerulesep}
7 & $\frac{1193745058447655963}{11989554297204369378}$ & $\frac{249900145094950907}{41963440040215292823}$ & $\frac{3347923855}{31727676806}$\tabularnewline[\doublerulesep]
\noalign{\vskip\doublerulesep}
\hline 
\noalign{\vskip\doublerulesep}
8 & $\frac{8835071648423645732519}{89051351248492234913674}$ & $\frac{1619796777034753048635}{356205404993968939654696}$ & $\frac{208657158071}{2010948047656}$\tabularnewline[\doublerulesep]
\noalign{\vskip\doublerulesep}
\hline 
\noalign{\vskip\doublerulesep}
9 & $\frac{3973328570636277936805618733}{40145601162806730995798838798}$ & $\frac{71993860817379312406691717}{20072800581403365497899419399}$ & $\frac{77376513420899}{754454218879206}$\tabularnewline[\doublerulesep]
\noalign{\vskip\doublerulesep}
\hline 
\noalign{\vskip\doublerulesep}
10 & $\frac{301536401029756814793984861993821417}{3051937062498858520392837449661769750}$ & $\frac{8855414478157869352976950222751601}{3051937062498858520392837449661769750}$ & $\frac{720930717976908431}{7088573434474257625}$\tabularnewline[\doublerulesep]
\hline 
\noalign{\vskip\doublerulesep}
\end{tabular}

\caption{Exact densities of critical percolation clusters on the lattice in
standard orientation rolled into a cylinder \label{tab:Exact-densities}}
\end{table}
As we noted, this $\alpha=\pi/4$ case corresponds to the percolation
on the lattice of standard orientation. Indeed the values of $\nu(l,1,1)$
 reproduce the values obtained in \cite{ChangShrock2021} for $l=2,\dots,5$,
where they appear as particular values of rational functions of the
open bond probability, when the value of the latter is set to $1/2$.
The other values of $\nu(l,1,1)$  are also confirmed
by the results of numerical simulations \cite{ZiffPrivate}. One can see that the length
of the rational numbers obtained grows quickly with $l$, which reveals
their combinatorial complexity. Remarkably, the lengths of numbers
representing $\nu(l,m,n)$ is significantly smaller than those of
both $\nu_{\mathrm{c}}(l,m,n)$ and $\nu_{\mathrm{nc}}(l,m,n)$ despite the former is the sum of the latters.  
It would be interesting to understand the origin of this fact. 
In \ref{appendix} we show a few more examples of the exact
densities at other values of $m,n$ and $l$. One can see that the
length of the numbers also grows drastically with values of $m$ and
$n$ quickly reaching the limit of the page width.

\begin{figure}
\centering{}\includegraphics[width=0.5\textwidth]{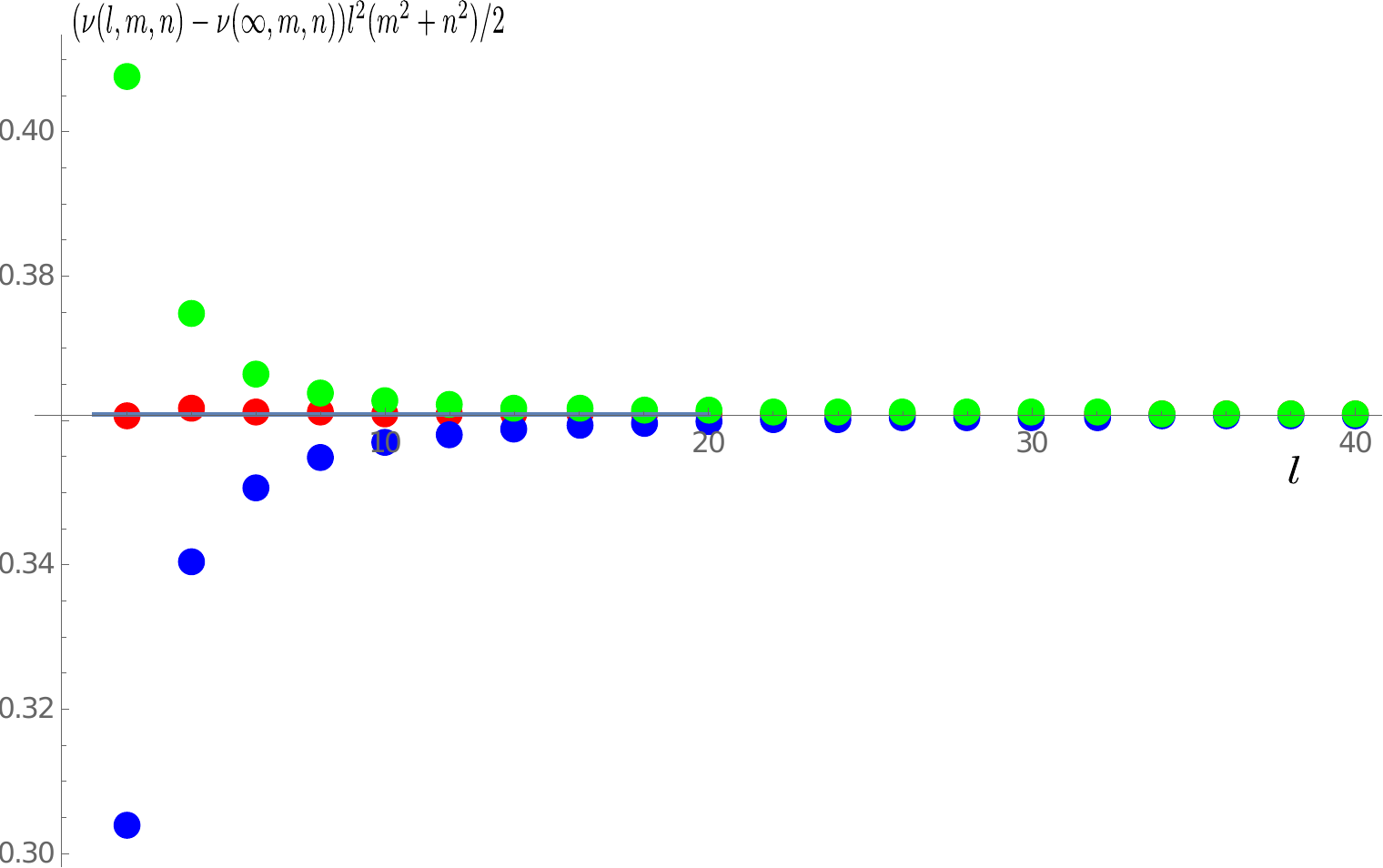}\caption{Convergence of the scaled difference $\left(\nu(l,m,n)-\nu(\infty,m,n)\right)l^{2}\left(m^{2}+n^{2}\right)/2$
to the limiting value $5\sqrt{3}/24\simeq0.360844$ for $(m,n)=(0,1)$
-- blue dots, $(m,n)=(1,1)$ -- green dots and $(m,n)=(1,2)$ --
red dots. \label{fig:Convergence}}
\end{figure}

It is much more informative to study the numerical values of the densities
obtained. With moderate computer resource the above procedure can
be performed with floating point calculations with precision to hundreds
of decimal digits for circumferences of a cylinder to hundreds. Then,
one can observe how the densities $\nu_{\mathrm{c}}(l,m,n)$ and $\nu(l,m,n)$
converge to its thermodynamic value
\[
\nu_{\mathrm{c}}(\infty,m,n)=\nu(\infty,m,n)=\frac{3\sqrt{3}-5}{2}\simeq0.098076,
\]
established in \cite{TL,ZiffFinchAdamchik}. The rate of convergence
is defined by the leading finite size corrections, the universality
of which provides the manifestation of the conformal invariance of
the model. For the total density of critical percolation clusters
the CFT based correction was predicted in  \cite{KlebanZiff} to be 
\[
\nu(l,1,1)-\nu(\infty,1,1)\simeq\frac{5\sqrt{3}}{24}\frac{1}{l^{2}}
\]
for the standard orientation of the lattice, corresponding to $(m,n)=(1,1)$
in our language. The universality suggests that for the rotated lattice
the dependence on the parameters $m$ and $n$ will enter only via
the length rescaling, 
\[
l\to l\sqrt{\left(m^{2}+n^{2}\right)/2}.
\]
Indeed convergence of the difference multiplied by the square of the
scaled circumference to the value of the coefficient is clear and
quick, see fig. \ref{fig:Convergence}.

In the same way one can study the next finite size corrections, which
though are not expected to be universal, still may contain a sort
of universal amplitudes. An example was given in \cite{KlebanZiff},
where the quantity
\begin{equation}
l^{4}\left(\nu(l,1,1)-\nu(\infty,1,1)-\frac{5\sqrt{3}}{24}\frac{1}{l^{2}}\right)\simeq a+\frac{b}{l^{2}}\label{eq: corrections}
\end{equation}
 was studied. The coefficients were estimated to
  $a=0.180$ and $b=0.69$.
Our estimate based on the fit of the scaled difference calculated
for the circumferences up to $l=100$ shows 
\begin{equation}
a=0.1804(2), b=0.47(6). \label{eq: a,b estimate}
\end{equation}
\begin{figure}
\centering{}\includegraphics[width=0.5\textwidth]{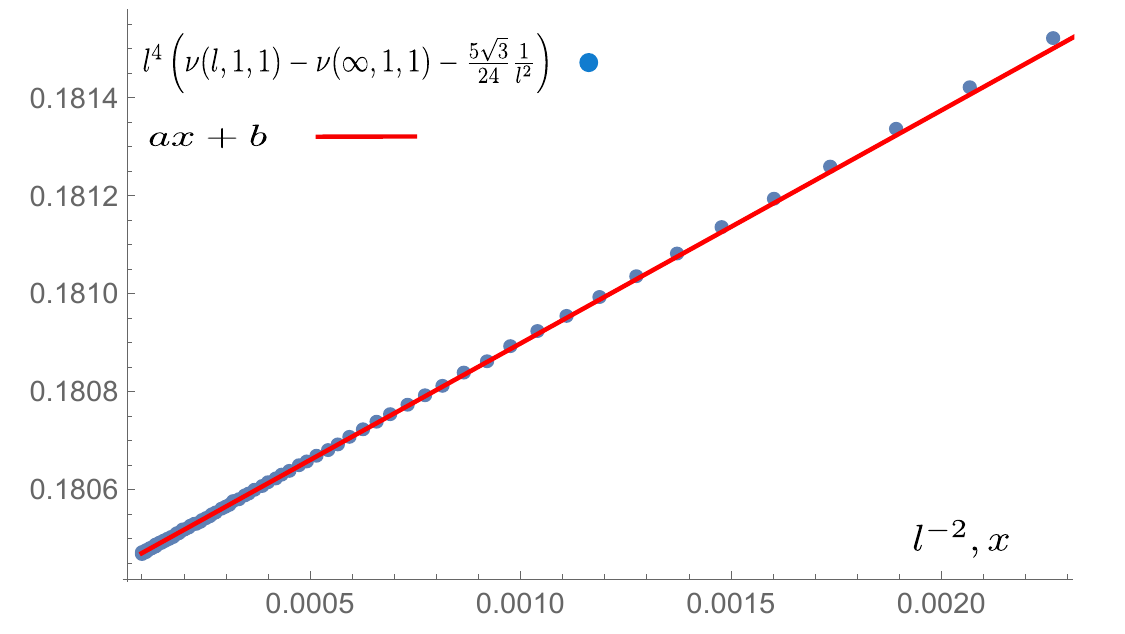}\caption{Comparison of the corrections to the densities (\ref{eq: corrections})
calculated at $20\protect\leq l\protect\leq100$ with the linear fit
with parameters $a=0.1804$ and $b=0.47.$}
\end{figure}

\begin{figure}
\centering{}\includegraphics[width=0.5\textwidth]{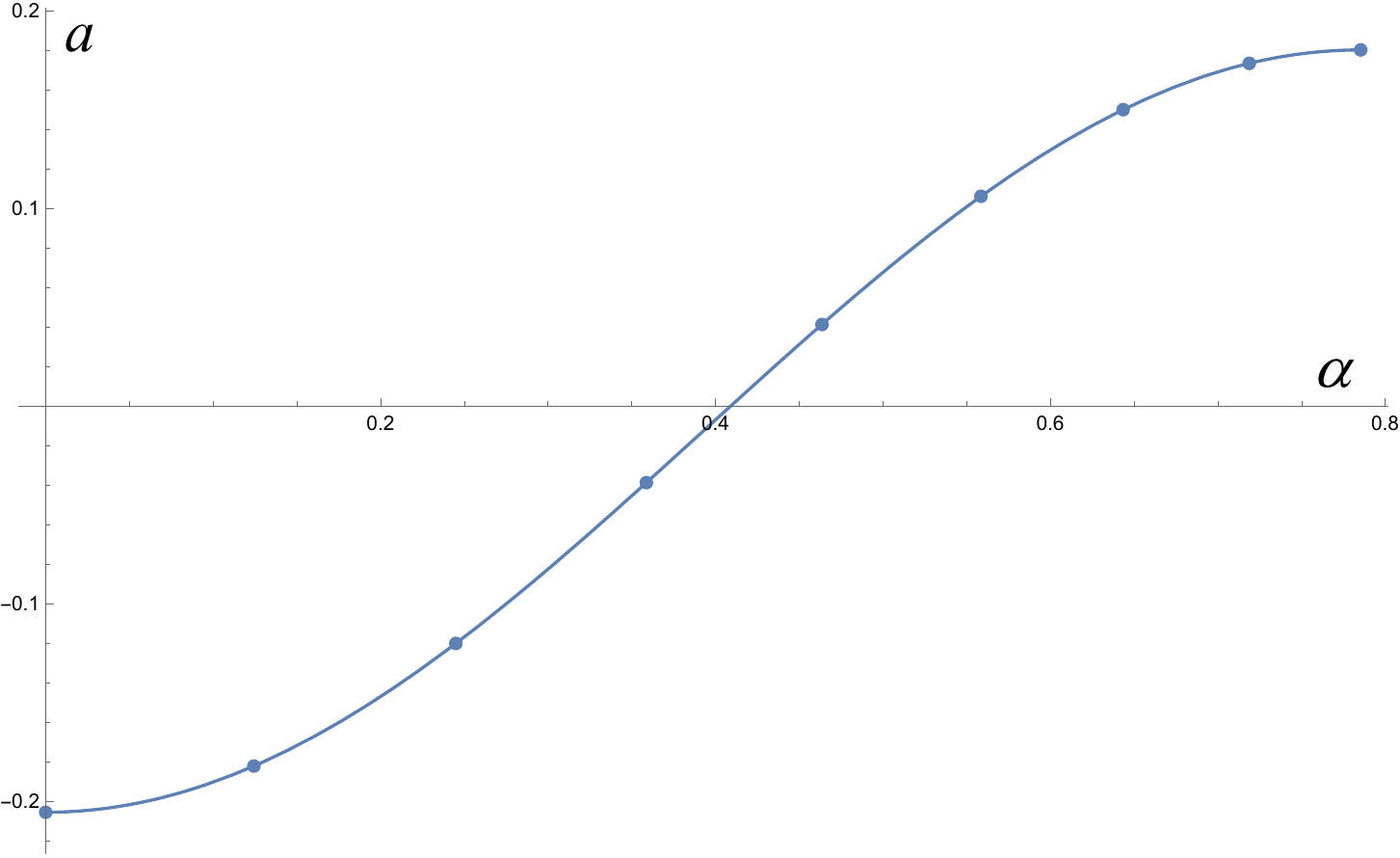}\caption{The values of scaled  correction coefficient $a(\alpha)$ at ten values of the  rotation angle $\alpha$ and its  fit by $c_{1}+c_2\cos4\alpha$
with $c_{1}=-0.0125,c_{2}=-0.192$.\label{fig: a_plot}}
\end{figure}

\begin{figure}
\centering{}\includegraphics[width=0.5\textwidth]{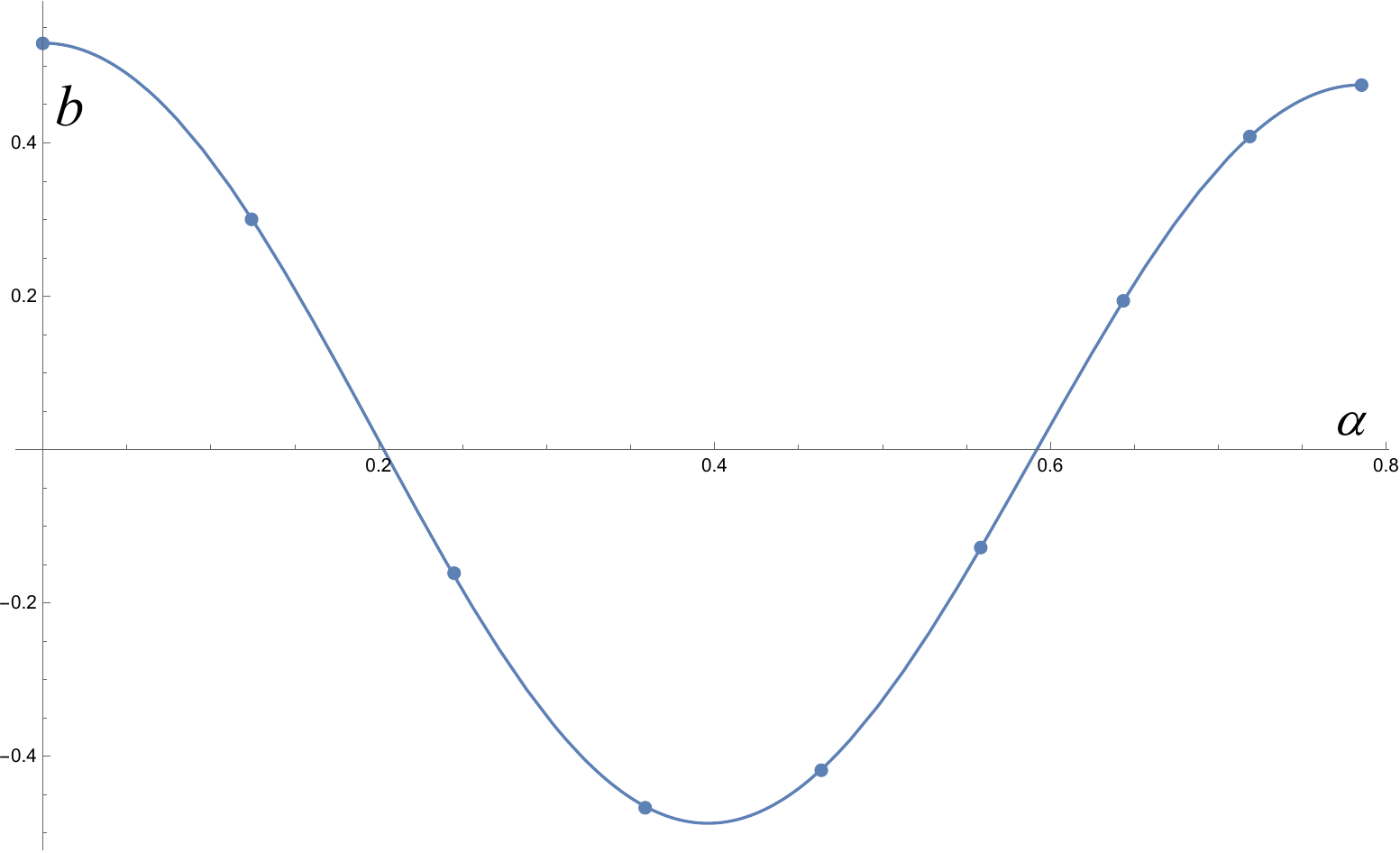}\caption{The values of scaled  correction coefficient $b(\alpha)$ at ten values of the  rotation angle $\alpha$ and its  fit by $c_{3}+c_4\cos(4\alpha)+c_5\cos(8\alpha)$
with $c_{3}=0.00760,c_4=0.0273,c_5=0.495$. \label{fig: b_plot}}
\end{figure}

It is interesting to see how the finite size corrections depend on
the angle. This behavior may shed light on irrelevant operators of
the theory responsible for the breaking of conformal invariance by
the lattice. We perform similar  at ten  values of $(n,m)$ and
study the dependence of the coefficients 
\[
a(\alpha)=\lim_{l\to\infty}\left(\frac{l^{2}\left(m^{2}+n^{2}\right)}{2}\right)^{2}\left(\nu(l,m,n)-\nu(\infty,m,n)-\frac{5\sqrt{3}}{12l^{2}\left(m^{2}+n^{2}\right)}\right)
\]
 and 
\[
\!\!\!\!\!\!\!\!\!\!\!\!\!\!\!\!\!\!\!\!\!\!\!\!\!\!b(\alpha)=\lim_{l\to\infty}\left(\frac{l^{2}\left(m^{2}+n^{2}\right)}{2}\right)^{3}\left(\nu(l,m,n)-\nu(\infty,m,n)-\frac{5\sqrt{3}}{12l^{2}\left(m^{2}+n^{2}\right)}-\frac{4a(\alpha)}{l^{4}\left(m^{2}+n^{2}\right)^{2}}\right)
\]
 on the angle $\alpha=\arctan(m/n)$. The results are presented in
figs.~\ref{fig: a_plot},\ref{fig: b_plot}.
They have clear periodic structure and are well fit with
\begin{eqnarray*}
a(\alpha) & \simeq & c_{1}+c_{2}\cos(4\alpha),\\
b(\alpha) & \simeq & c_{3}+c_4\cos(4\alpha)+c_5\cos(8\alpha),
\end{eqnarray*}
with $c_{1}=-0.0125,c_{2}=-0.192$ and $c_{3}=0.00760,c_4=0.0273,c_5=0.495$. 
The quality of the fit can  be tested  on  the exact values of expansion coefficients of  $\nu(l,0,1)$ obtained from asymptotic expansion   
\begin{equation}
\hspace{-1.5cm}\nu(l,0,1)=\left(\frac{3 \sqrt{3}}{2}-\frac{5}{2}\right)+\frac{5}{8 \sqrt{3} }\left(\frac{2}{l^2}\right)-\frac{205}{576 \sqrt{3}} \left(\frac{2}{l^2}\right)^2+\frac{19025} {20736 \sqrt{3} }\left(\frac{2}{l^2}\right)^3+O\left(\frac{1}{l^8}\right)
\end{equation}
of the exact result in \cite{Povolotsky2021}.
One can see that the the values of constants $c_1,\dots,c_5$ obtained from the fits satisfy relations 
\begin{eqnarray}
	a(0)=c_1+c_2= -0.2055\simeq -\frac{205}{576 \sqrt{3}},\\
	b(0)=c_3+c_4+c_5=0.5298\simeq \frac{19025} {20736 \sqrt{3} }
\end{eqnarray} 
with accuracy to three decimal places. The values of  $a(\pi/4)=c_1-c_2,b(\pi(4))=c_3-c_4+c_5$ reproduce (\ref{eq: a,b estimate}) within the same accuracy.
As explained in \cite{Cardy1988}, the 
trigonometric functions  of angles divisible by $4\alpha$ is a natural manifestation of
breaking  the rotational symmetry of the
conformally invariant theory  by the square lattice that results in appearance of  operators whose  conformal spin is multiple of four, see also \cite{CJV2017,TDJ2019} for similar effects. 
It is interesting whether the values of constants $c_1,\dots,c_5$ can be explained in the framework of CFT. This analysis can be considered as
a preliminary step in systematic studies of the finite size corrections
to cluster densities, and clearing up their conformal meaning. 

For more detailed  study of the finite size corrections one needs to perform
the systematic asymptotic analysis of the solution of T-Q equation. This hopefully 
can be done with the non-linear integral equation technique developed in \cite{KlumperBatchelorPearce1991,KlumperWehnerZittartz} and applied in \cite{Fujimoto1996} to a proof of the conformal invariant form of the sub-leading finite size correction to free energy of the six-vertex model on the rotated lattice. Whether this technique is suitable for systematic analysis of the next to sub-leading corrections to the derivatives of the free energy at the stochastic point is the matter for further investigation.     

\ack I am grateful to Robert Ziff for encouraging discussion and informing me about the available  
data on critical percolation cluster densities. I am grateful to Robert Shrock 
for attracting my attention to  results of articles \cite{ChangShrock2004,ChangShrock2021}. 
  The work is supported by Russian Foundation of Basic Research under grant 20-51-12005.\\
  
\bibliography{O1_rotated_revised}

\appendix\newpage{}

\section{Exact densities\label{appendix}}

\subsection*{$m=1,n=2$}
\begin{table}[H]
\begin{tabular}{|l|l|}
\hline 
$l$ & $\nu_{\mathrm{c}}(l,1,2)$\tabularnewline
\hline 
\hline 
\noalign{\vskip\doublerulesep}
$2$ & $\frac{457}{4346}$\tabularnewline[\doublerulesep]
\noalign{\vskip\doublerulesep}
\hline 
\noalign{\vskip\doublerulesep}
$4$ & $\frac{2156027742167}{21586801226362}$\tabularnewline[\doublerulesep]
\noalign{\vskip\doublerulesep}
\hline 
\noalign{\vskip\doublerulesep}
$6$ & $\frac{431777998328921413810430569}{4366789318056412396314592982}$\tabularnewline[\doublerulesep]
\noalign{\vskip\doublerulesep}
\hline 
\noalign{\vskip\doublerulesep}
$8$ & $\frac{786294430244725543429301774315042465275143236939999}{7980483750425370301607306038647768778325881306171354}$\tabularnewline[\doublerulesep]
\noalign{\vskip\doublerulesep}
\hline 
\noalign{\vskip\doublerulesep}
$10$ & $\frac{3063214625120684629220276976212585297353537811730106923006124169424851246746977611251}{31141354303881719194901531792574953430390711629201099942966304916168344436205188871378}$\tabularnewline[\doublerulesep]
\hline 
\noalign{\vskip\doublerulesep}
\end{tabular}

\end{table}

\begin{table}[H]
\begin{tabular}{|l|l|}
\hline 
$l$ & $\nu_{\mathrm{nc}}(l,1,2)$\tabularnewline
\hline 
\hline 
\noalign{\vskip\doublerulesep}
$2$ & $\frac{63}{2173}$\tabularnewline[\doublerulesep]
\noalign{\vskip\doublerulesep}
\hline 
\noalign{\vskip\doublerulesep}
$4$ & $\frac{78131053098}{10793400613181}$\tabularnewline[\doublerulesep]
\noalign{\vskip\doublerulesep}
\hline 
\noalign{\vskip\doublerulesep}
$6$ & $\frac{7013973470238361101359808}{2183394659028206198157296491}$\tabularnewline[\doublerulesep]
\noalign{\vskip\doublerulesep}
\hline 
\noalign{\vskip\doublerulesep}
$8$ & $\frac{7205724473170424432979362883354946667863166859216}{3990241875212685150803653019323884389162940653085677}$\tabularnewline[\doublerulesep]
\noalign{\vskip\doublerulesep}
\hline 
\noalign{\vskip\doublerulesep}
$10$ & $\frac{89949785674341668790750642180399072760356918123516183594046343010414953902267663104}{77853385759704297987253829481437383575976779073002749857415762290420861090512972178445}$\tabularnewline[\doublerulesep]
\hline 
\noalign{\vskip\doublerulesep}
\end{tabular}

\end{table}

\begin{table}[H]
\begin{tabular}{|l|l|}
\hline 
$l$ & $\nu(l,1,2)$\tabularnewline
\hline 
\hline 
\noalign{\vskip\doublerulesep}
$2$ & $\frac{11}{82}$\tabularnewline[\doublerulesep]
\noalign{\vskip\doublerulesep}
\hline 
\noalign{\vskip\doublerulesep}
$4$ & $\frac{275693}{2573782}$\tabularnewline[\doublerulesep]
\noalign{\vskip\doublerulesep}
\hline 
\noalign{\vskip\doublerulesep}
$6$ & $\frac{22090069080005}{216378176911486}$\tabularnewline[\doublerulesep]
\noalign{\vskip\doublerulesep}
\hline 
\noalign{\vskip\doublerulesep}
$8$ & $\frac{8846218291444135690643021}{88168581201012456573730414}$\tabularnewline[\doublerulesep]
\noalign{\vskip\doublerulesep}
\hline 
\noalign{\vskip\doublerulesep}
$10$ & $\frac{12385236427151367035550969161665589631653003}{124449443496736607229791988937036731786613090}$\tabularnewline[\doublerulesep]
\hline 
\noalign{\vskip\doublerulesep}
\end{tabular}

\end{table}

\subsection*{
$m=1,n=4$}

\begin{table}[H]
\begin{tabular}{|l|l|}
\hline 
$l$ & $\nu_{\mathrm{c}}(l,1,4)$\tabularnewline
\hline 
\hline 
\noalign{\vskip\doublerulesep}
$2$ & $\frac{3417329}{34119046}$\tabularnewline[\doublerulesep]
\noalign{\vskip\doublerulesep}
\hline 
\noalign{\vskip\doublerulesep}
$4$ & $\frac{4131823530849759693213650061797}{41903044260219093187283072814910}$\tabularnewline[\doublerulesep]
\noalign{\vskip\doublerulesep}
\hline 
\noalign{\vskip\doublerulesep}
$6$ & $\frac{52354510386464168141080291945089720530592239055392948076663254479940653}{532536631210089043138388552997130780273523772549104271499136454928074734}$\tabularnewline[\doublerulesep]
\hline 
\noalign{\vskip\doublerulesep}
\end{tabular}

\end{table}

\begin{table}[H]
\begin{tabular}{|l|l|}
\hline 
$l$ & $\nu_{\mathrm{nc}}(l,1,4)$\tabularnewline
\hline 
\hline 
\noalign{\vskip\doublerulesep}
$2$ & $\frac{143712}{17059523}$\tabularnewline[\doublerulesep]
\noalign{\vskip\doublerulesep}
\hline 
\noalign{\vskip\doublerulesep}
$4$ & $\frac{44386921318627731749339096064}{20951522130109546593641536407455}$\tabularnewline[\doublerulesep]
\noalign{\vskip\doublerulesep}
\hline 
\noalign{\vskip\doublerulesep}
$6$ & $\frac{52354510386464168141080291945089720530592239055392948076663254479940653}{532536631210089043138388552997130780273523772549104271499136454928074734}$\tabularnewline[\doublerulesep]
\hline 
\noalign{\vskip\doublerulesep}
\end{tabular}

\end{table}

\begin{table}[H]
\begin{tabular}{|l|l|}
\hline 
$l$ & $\nu(l,1,4)$\tabularnewline
\hline 
\hline 
\noalign{\vskip\doublerulesep}
$2$ & $\frac{1007}{9274}$\tabularnewline[\doublerulesep]
\noalign{\vskip\doublerulesep}
\hline 
\noalign{\vskip\doublerulesep}
$4$ & $\frac{950775017326855}{9439509175366346}$\tabularnewline[\doublerulesep]
\noalign{\vskip\doublerulesep}
\hline 
\noalign{\vskip\doublerulesep}
$6$ & $\frac{45931697658977271960510394658373737}{462768523437381762199714804256505814}$\tabularnewline[\doublerulesep]
\noalign{\vskip\doublerulesep}
\hline 
\noalign{\vskip\doublerulesep}
\end{tabular}

\end{table}

\subsection*{$
m=3,n=4
$}
\begin{table}[H]
\begin{tabular}{|l|l|}
\hline 
$l$ & $\nu_{\mathrm{c}}(l,3,4)$\tabularnewline
\hline 
\hline 
\noalign{\vskip\doublerulesep}
$2$ & $\frac{387768065791915313}{3895953850182322426}$\tabularnewline[\doublerulesep]
\noalign{\vskip\doublerulesep}
\hline 
\noalign{\vskip\doublerulesep}
$4$ & $\frac{44195116406974440581081566067289510598931484194160698978934613738812968820065999}{448965092769002158674204818544686846143985958937884255868928657300610488746952074}$\tabularnewline[\doublerulesep]
\hline 
\noalign{\vskip\doublerulesep}
\end{tabular}

\end{table}

\begin{table}[H]
\begin{tabular}{|l|l|}
\hline 
$l$ & $\nu_{\mathrm{nc}}(l,3,4)$\tabularnewline
\hline 
\hline 
\noalign{\vskip\doublerulesep}
$2$ & $\frac{11343951399731931}{1947976925091161213}$\tabularnewline[\doublerulesep]
\noalign{\vskip\doublerulesep}
\hline 
\noalign{\vskip\doublerulesep}
$4$ & $\frac{324703379991987605772074955808303082817396322520577832155520445581250214007026}{224482546384501079337102409272343423071992979468942127934464328650305244373476037}$\tabularnewline[\doublerulesep]
\hline 
\noalign{\vskip\doublerulesep}
\end{tabular}
\end{table}

\begin{table}[H]
\begin{tabular}{|l|l|}
\hline 
$l$ & $\nu(l,3,4)$\tabularnewline
\hline 
\hline 
\noalign{\vskip\doublerulesep}
$2$ & $\frac{505753025}{4800491638}$\tabularnewline[\doublerulesep]
\noalign{\vskip\doublerulesep}
\hline 
\noalign{\vskip\doublerulesep}
$4$ & $\frac{1025205092866657199069018079406719010967}{10263935640757775218890055689049451622658}$\tabularnewline[\doublerulesep]
\noalign{\vskip\doublerulesep}
\hline 
\noalign{\vskip\doublerulesep}
\end{tabular}
\end{table}

\end{document}